\documentclass{mn2e}
\usepackage{times}
\input{psfig.sty}
\onecolumn
\newif\ifAMStwofonts
\newcommand{\pderiv}[2]{\frac{\partial #1}{\partial #2}}

\title{An explanation for long flares from extragalactic globular cluster X-ray sources}

\author[Maccarone et al.] {Thomas J. Maccarone\\
Astronomical Institute ``Anton Pannekoek'', University of Amsterdam,
Kruislaan 403, 1098 SJ, Amsterdam, The Netherlands\\ and\\ School of Physics and Astronomy, University of Southampton, Southampton, Hampshire, SO17 1BJ} 

\date{}

\begin{document}

\maketitle

\label{firstpage}

\def\simlt{\mathrel{\rlap{\lower 3pt\hbox{$\sim$}}
        \raise 2.0pt\hbox{$<$}}}
\def\simgt{\mathrel{\rlap{\lower 3pt\hbox{$\sim$}}
        \raise 2.0pt\hbox{$>$}}}

\input epsf

\begin{abstract}
Repeatedly flaring X-ray binaries have recently been discovered in
NGC~4697 by Sivakoff and collaborators.  We show that these flares can
be explained as the result of eccentric binaries in globular clusters
which accrete more rapidly at periastron than during the rest of the
binary orbit.  We show that theoretical timescales for producing
eccentricities and circularising the binaries are consistent with what
is needed to produce the observed population of flaring sources,
although the circularisation timescales are highly uncertain on both
observational and theoretical grounds.  This model makes two clear
theoretical predictions (1) the flares should be seen to be strictly
periodic if adequate sampling is provided, and that periodicity should
be of approximately 15 hours (2) this class of flaring behaviour
should be seen only in globular cluster sources, and predominantly in
the densest globular clusters. We also test the model for producing
eccentricities through fly-by's of a third star near the binary in a
globular cluster against a much larger database of millisecond pulsar
observations than has been used in past work, and find that the
theoretical cross sections for producing eccentricity in binaries are
in reasonable agreement with most of the data, provided that the
pulsar ages are about $4\times10^9$ years.
\end{abstract}


\section{Introduction}

Extragalactic globular clusters span a much wider range of age and
metallicity than do Galactic globular clusters.  They also contain
many more globular clusters; M~87, for example, contains about 10,000
globular clusters -- two orders of magnitude more than the Milky Way
does.  As a result, rare events in globular clusters may be observable
in other galaxies, but not in our own.  At the present time,
extragalactic globular clusters are too dense for the individual stars
to be resolved, complicating searches for unusual stars that might be
the results of interactions in these dense environments.  It is likely
though, that most globular clusters contain at most one bright X-ray
binary, and these X-ray binaries can be observed to distances of tens
of Megaparsecs.  We could thus have some hope of finding signatures of
unusual properties of the X-ray binaries that manifest themselves in
the X-rays.

Identifying unusual X-ray binaries in globular clusters is more easily
done in elliptical galaxies in particular than in our own Galaxy.
This is partly because the metal rich globular clusters found more
prevalently in elliptical galaxies than in spirals.  Metal rich
clusters are about three times as likely to contain an X-ray binary as
metal poor clusters (Kundu, Maccarone \& Zepf 2002), and this may
contribute to the fact that ellipticals typically have 4\% of their
globular clusters with bright X-ray sources, while for spiral
galaxies, this number is typically about 1-3\% (Maccarone, Kundu \&
Zepf 2003).

In this paper, we show that recently discovered flaring X-ray binaries
in NGC~4697 (Sivakoff, Sarazin \& Jordan 2005 -- SSJ05) may be
presenting evidence of eccentric X-ray binaries in these globular
clusters.  We show that the observational data are consistent with the
expectations for such eccentric binaries, and that it is reasonable
for a few percent of the X-ray binaries to be eccentric enough for
their accretion rates to be affected at periastron passage.  We
conclude that it is likely that such sources have not been found in
the Milky Way because of small number statistics, and note that an
eccentric binary model for a class of flaring behaviour makes two
clear predictions: (1) that the flaring should be strictly periodic
and (2) that the flaring sources in this class should be seen only in
globular clusters, and preferentially in the densest globular
clusters.

\section{The observational data}
We briefly summarise the results of SSJ05.  In that paper, they found
evidence for strong flaring from three X-ray binaries in NGC~4697.
The sample contained 157 X-ray sources, about 90\% of which should be
X-ray binaries, so $\sim2\%$ of the X-ray binaries showed this strong
flaring behaviour.  In two of the systems, the flaring had similar
characteristics: frequent (i.e. roughly once per 10-15 hours) flares
with durations of about 1000 seconds, luminosities of about
$5\times10^{38}$ ergs/sec (0.5-10 keV) and total flare energies of
about $5\times10^{41}$ ergs.  The estimate of the frequency of the
flares in these sources is quite rough, due to the relatively small
number of observations and the sparse sampling of the data.  SSJ05
note that the two most closely spaced flares in a single source
occurred 11 days apart, and that there could be an additional
population of sources which flare much more rarely, and whose flares
simply did not occur within the observation windows.  Thus, while the
data are consistent with the idea that only a small number of sources
flare, and that they flare with a frequency of about once per 15
hours, this is not a unique interpretation of the data.  The third
flaring system showed much shorter (duration of about 1 minute), much
brighter flares (luminosities of about $5\times10^{39}$ ergs/sec).
The third flare was suggested by SSJ05 to be flaring activity from a
``microblazar,'' an X-ray binary whose jet is pointed at the observer
(see e.g. Mirabel \& Rodriguez 1999; K\"ording, Falcke \& Markoff
2002).

\section{Why superbursts cannot explain the data}
The only possible explanation put forth by SSJ05 was that the two
longer flares were superbursts.  It was their intention that the
reader would infer that they had dismissed the possibility
(G. Sivakoff \& C. Sarazin, private communications) due to the
observational differences with Galactic superbursters discussed in
their paper.  We repeat here their discussion of what are the
phenomenological differences between superbursts and the flares seen
in NGC~4697, and place these differences in their theoretical context
in order to make it more clear to readers not versed in burst theory
that the observed differences are truly fundamental ones, and not
differences that could be accounted for by tweaking parameter values
such as accretion rate or the chemical composition of the accreted
material.

Type I X-ray bursts as a broad class are generally explained as the
runaway nuclear burning of piled-up accreted material in a layer on
the surface of the accreting neutron star (for a review, see Lewin et
al. 1993).  The luminosities of these bursts are found to be at the
Eddington limit or slightly below (Kuulkers et al. 2003).  The decay
timescale distribution of Type I X-ray bursts shows a bimodal
distribution, with peaks at about 5 and 15 seconds (Kuulkers 2004).

Superbursts are a special class of Type I X-ray bursts, with much
longer durations, typically about 3000--10000 seconds.  Despite
superbursts' producing many more photons than classical Type I X-ray
bursts, they were discovered decades after the first Type I X-ray
bursts (Cornelisse et al. 2000), highlighting their rarity.  It is
generally believed that the same basic mechanism (i.e. runaway nuclear
burning) is taking place in superbursts as in normal Type I X-ray
bursts, but that instead of the bursts being powered by runaway
thermonuclear burning of hydrogen and helium, they are believed to be
powered by burning of carbon (Cumming \& Bildsten 2001), or perhaps by
a more complicated mechanism in which hydrogen atoms capture electrons
to become neutrons, and then neutron capture by heavier atoms deeper
in the atmosphere of the neutron star occurs (Kuulkers et al. 2002).

In only one case was a superburst found to have a peak luminosity in
excess of the Eddington limit for solar composition material onto a
1.4 $M_\odot$ neutron star, that of the superburst from 4U~1820-30
(Strohmayer \& Brown 2002).  It should be noted though, that in this
case, the bursting system is known to be an ultracompact X-ray binary,
where the accreted material is predominantly helium, and where the
effective Eddington limit will thus be about twice as large as for
accretion of hydrogen.  There is thus no evidence for superbursts'
peak luminosities being more than about 10\% above the Eddington
limit.  Additionally, since Type I bursts are generally thought to be
spherical in nature (i.e. nuclear burning over the whole surface of
the neutron star), it is expected that the Eddington limit should
hold for these events, even if it might be possible to exceed the
Eddington limit in the case of non-spherical accretion.

Several authors have also attempted to estimate the recurrence
timescale of superbursts, based on both observational and theoretical
grounds.  Observationally, the recurrence timescale appears to be
about one to two years (Wijnands 2001; Kuulkers 2002; in 't Zand et
al. 2003).  Theoretical models predict similar (Cumming 2003) or
somewhat longer values (Strohmayer \& Brown 2002).  For superbursts to
be seen from the same source as frequently as the flaring sources
discovered by SSJ05 would thus be indicating new nuclear physics, and
not just a new range of parameter space being filled by accreting
sources.  For a more detailed review of the properties of superbursts,
we refer the reader to Kuulkers (2004).

The properties of the Galactic superbursts thus present a few clear
differences relative to the properties of the flaring events seen by
SSJ05.  The most important is the recurrence time difference.  The two
sources suggested by SSJ05 to be candidate superbursters each showed a
flare in 3 of the 5 observations of NGC~4697, which had durations of
about 40 kiloseconds each.  This gives a flare rate of about one per
70000 seconds, or about 1000 times as often as the Galactic
superbursts appear to occur.  As the superburst recurrence timescale
is set largely by nuclear physics, it is highly unlikely that simply
tweaking parameter values could increase the superburst rates to the
levels seen in the NGC~4697 sources.

The luminosities of four of the six events claimed to be superbursts
are above the Eddington limit even for a helium-accreting neutron
star, and the event durations of 5 out of the six candidates are less
than 2000 seconds.  Furthermore, the conversion from count rate to
luminosity was done using a $\Gamma=1.4$ power law, which yields a
luminosity about 2.7 times lower than a more realistic spectral model
for a superburst (SSJ05).  This means that most of the flares are at
least a factor of 5 times brighter than any known Galactic superburst,
further weakening the cases that these events are truly superbursts.
While the measured event durations are also inconsistent with the
observations of Galactic superbursts, this quantity is highly
susceptible to measurement error for the extragalactic flares, since
there are relatively few photons from which to estimate the light
curve shape of the burst.

\section{Eccentric binaries}
An alternative way to produce multiple bright flares in a single X-ray
binary is to have a frequently occurring type of event that boosts the
actual accretion rate.  The most natural such type of event would be a
periastron passage in an eccentric binary.  One can consider the case
of the Galactic X-ray binary Circinus X-1 as an example of such a
system.  This object has a 16.6 day orbital period, with strong flares
in the X-rays seen at each periastron passage (e.g. Kaluzienski et
al. 1976; Clarkson, Charles \& Onyett 2004).

On the other hand, the eccentricity of Cir X-1 is quite large, with
estimates in the 0.80-0.95 range (e.g. Tauris et al. 1999).  There are
no other low mass X-ray binaries with known eccentricities, so there
is no clear empirical evidence about what to expect for the effects on
the accretion rate of an eccentric low mass X-ray binary with small,
but non-zero, eccentricity.  However, theoretical predictions have
been made.  Hut \& Paczynski (1984) considered the cases of isothermal
flows and polytropic flows and found that:
\begin{equation}
\pderiv{{\rm ln} \dot{M}}{{\rm ln} (R_2/R_{L2})} \approx 2\times10^4,
\end{equation}
where $\dot{M}$ is the mass accretion rate, $R_2$ is the radius of the
mass donor star, and $R_{L2}$ is the distance from the center of the
mass donor star to the second Lagrangian point (i.e. the size of the
Roche lobe of the mass donor).  Slightly larger values are found for
isothermal flows, and slightly smaller values for polytropic flows.
Therefore, if one changes the eccentricity from zero to 0.001 while
leaving the semi-major axis unchanged, that is enough to increase the
mass transfer rate by a factor of 100 (Hut \& Paczynski 1984).

The next question is whether it is feasible for the changes in the
mass accretion rate to manifest themselves as periodic behaviour.  The
relevant timescale is then the viscous timescale from the outer part
of the accretion disk.  The viscous timescale, $t_{visc}$, can be
defined in terms of the dynamical timescale, $t_{dyn}$, such that:
\begin{equation}
t_{visc}=(H/R)^{-2} \alpha^{-1} t_{dyn},
\end{equation}
where $H$ is the scale height of the disk, $R$ is the radius of the
location in the disk where the viscous timescale is being computed,
and $\alpha$ is the dimensionless viscosity parameter.  Since the
accretion rates we are considering are quite high, the accretion is
likely to proceed through a geometrically thick accretion disk, with
$H/R\sim1$, and $\alpha$ likely to be relatively large for large
accretion rates (Blaes 1987; Abramowicz et al. 1988).  The
circularisation radius is typically about 1/7 of the orbital
separation, yielding a dynamical timescale at the circularisation
radius of about 1/20 of the orbital period.  A value of $\alpha$ of
about 0.3 or more would then give timescales for the flare events that
are consistent with the measurements of SSJ05, assuming that the
orbital periods of the binaries are about 15 hours -- a bit longer
than the observations used to find the flare -- as would be expected
by the finding that the flaring sources each showed flares in 3 of the
5 observations.

That the two flaring sources are seen in globular clusters bolsters
the idea that they might be eccentric.  Globular cluster X-ray
binaries are generally thought to be produced through dynamical
interactions of some kind -- tidal captures (Clark 1975; Fabian,
Pringle \& Rees 1975) or three-body or four-body exchange interactions
(e.g. Hills 1976; Fregeau et al. 2004).  These interactions are likely
to leave behind highly eccentric binaries, or in some cases, even to
leave behind products of stellar collisions.  Another possibility that
is less often discussed is that accretion takes place in systems
hardened by interactions with other stars or binaries (Krolik, Meiskin
\& Joss 1984).  These interactions will also perturb the binary orbits
so that the binaries become eccentric (Hut \& Paczynski 1984;
Rappaport, Putney \& Verbunt 1989; Phinney 1992; Rasio \& Heggie 1995;
Heggie \& Rasio 1996).

\section{Rate of perturbations to the eccentricity}
Two cases exist for producing eccentric X-ray binaries -- first that
the eccentricity is induced in the encounter forming the X-ray binary,
and secondly that the eccentricity is induced after formation.  Recent
numerical simulations do indicate that a large fraction of 3-body and
4-body interactions produce eccentricities well in excess of 0.1
(Fregeau et al. 2004). Many products of tidal capture (Di Stefano \&
Rappaport 1990) are also expected to be highly eccentric.  It is not
clear how stable such systems are; they probably undergo thermal
timescale mass transfer and have very short lives as X-ray sources
(Hut \& Paczynski 1984).  In the interests of making conservative
estimates of how many eccentric X-ray binaries there should be, we
will focus on the case where an initially circular binary has an
eccentricity induced by a third star passing near to it.  This most
likely corresponds to the real formation scenario of an eccentric, but
non-contact binary being formed by a stellar interaction,
circularising, and then coming into contact due to gravitational
radiation and stellar evolutionary effects.

We thus wish to find whether the duty cycle as an eccentric source is
large enough to account for the observed fraction of the objects
showing bright flares.  Let us consider the case where the
eccentricity enhances the mass transfer rate by a factor of about 10
(since the flares observed by SSJ05 are at a count rate of about 10
times the non-flaring level).  This requires an eccentricity of about
0.0001, though the exact value depends on whether the star is better
treated as a polytrope or as an isothermal flow at the inner
Lagrangian point (Hut \& Paczynski 1984).

We can estimate the probability that a given eccentricity will be
induced in an X-ray binary by using cross-sections for eccentricity
tabulated by various authors.  The most recent work on this topic is
that of Heggie \& Rasio (1996).  For the case where small
eccentricities are generated, they find that:
\begin{equation}
\sigma(\delta{e}>\delta{e_0})=4.62 \left(\frac{m_3|m_1-m_2|}{M_{12}^2}\right)^{2/5} \left(\frac{M_{12}}{M_{123}}\right)^{1/5}\frac{GM_{123}a}{V^2} \delta{e_0}^{-2/5}, 
\label{cross}
\end{equation}
where $\sigma(\delta{e}>\delta{e_0})$ is the cross-section for
producing an eccentricity increases greater than $\delta{e_0}$, $m_1$
is the accretor's mass, $m_2$ is the donor star's mass, $m_3$ is the
mass of the third body passing nearby the binary, $M_{12}$ is the
total mass of the binary system, $M_{123}$ is the sum of the masses of
all three stars, $G$ is the gravitational constant, $a$ is the
semi-major axis of the binary system, and $V$ is the relative velocity
of the centre of mass of the binary and the third star.

We then follow the work of Hut \& Paczynski (1984), and compute the
timescale for an interaction to occur, assuming the velocity
distribution of the stars follows is Maxwellian.  Hut \& Paczynski
(1984) did point out correctly both that the lowered Maxwellian using
in King models would be a better approximation to the real velocity
distributions of globular clusters, and that the difference between
the two distributions is small except in the very high velocity tail,
which contributes at a very small level to the cross section anyways.
Since it will be straightforward to compute an analytic expression for
the interaction timescale only from the Maxwellian distribution, we
therefore use the Maxwellian.

The timescale, $\tau$, for an interaction for a single binary is then
given by:
\begin{equation}
1/\tau = n <\sigma V> = n \left(\frac{2}{\pi}\right)^{1/2} \left(\frac{\mu}{kT}\right)^{3/2} \int_0^{\infty} {\rm exp} (-\mu v^2/2kT)\sigma(V)V^3dV,
\end{equation} 
where $n$ is the density of stars, $k$ is the Boltzmann constant,
$\mu=\frac{m_3(m_1+m_2)}{m_1+m_2+m_3}$ is the reduced mass of the
three star system, and $v_{th}=(3kT/M_\odot)^{1/2}$ defines the
temperature $T$, in terms of the thermal velocity $v_{th}$, which is
approximately the same as the velocity dispersion, as in the case of
the Maxwellian distribution, the globular cluster is approximated as a
gas of particles with masses of a solar mass and temperature T.  We
then re-write this expression, in terms of the velocity dispersion,
$\sigma_v$, rather than the ``temperature,'' in order to express the
results in terms of observables:
\begin{equation}
1/\tau = n <\sigma V> = n \left(\frac{2}{\pi}\right)^{1/2} \left(\frac{3\mu}{M_\odot \sigma_v^2}\right)^{3/2} \int_0^{\infty} exp [-3\mu v^2/(2M_\odot \sigma_v^2]\sigma(V)V^3dV,
\end{equation}
Substituting
the expression for the cross-section from Equation \ref{cross}, we
find:
\begin{equation}
\tau= 7\times10^{10} {\rm yr} \left(\frac{a}{0.01 {\rm AU}}\right)^{-1} \left(\frac{n}{10^5 {\rm pc}}\right)^{-1} \left(\frac{\sigma_{v}}{10 {\rm km/sec}}\right)\left(\frac{m_3|m_1-m_2|}{M_{12}^2}\right)^{-2/5} \left(\frac{M_{12}}{M_{123}}\right)^{-1/5} \left(\frac{\mu}{m_\odot}\right)^{-1/2} \left(\frac{M_{123}}{M_\odot}\right)^{-1} {e}^{2/5}
\end{equation}

This gives a timescale to produce the eccentricity of 0.0001 of about
$10^9$ years for parameter values of $m_1$=1.4 $M_\odot$,
$m_2=m_3=1.0M_\odot$, $n=5\times10^5$, $\sigma_v=10$ km/sec, and $P=$10
hrs, typical values in a dense globular cluster.  We note that this
formula breaks down for very close approaches, which can then yield
very large eccentricities.  For such close approaches, the
cross-section scales with the logarithm of the final eccentricity,
rather than as a power law function of the final eccentricity.  Such
cases will not be relevant for the low mass X-ray binaries we study
here.

We can also test the theoretical predictions of the timescale for
producing eccentricity by comparing with the millisecond pulsar
population of the Galaxy's globular clusters.  This was previously
done with inconclusive results by Rasio \& Heggie (1995), but that was
with only nine binary pulsars.  There are now 31 known binary pulsars
with good orbital data compiled in Camilo \& Rasio (2005).  For these
systems, we can find the expected $e$ given different assumptions
about the pulsar lifetime, and then plot the expected versus observed
values.  A few caveats will apply.  First, for pulsars which do not
have positions measured accurately, we will assume that the local
stellar density and velocity dispersion are the core density and core
velocity dispersion for the cluster, which will usually, but not
always, be a good approximation.  We will use a separate symbol to
identify these pulsars.  Secondly, we will compute the expected
eccentricity only in the power law regime, so the expected
eccentricities of the widest binaries will be underestimates.

We take the estimates of the globular cluster velocity dispersion from
Gnedin et al. (2002).  We assume the perturbing star will have a mass
of 1 $M_\odot$, and that the timescale for inducing a perturbation is
4 Gyr.  We also assume no stellar or orbital evolution over that 4
Gyr.  The central densities are taken from the Harris catalog (Harris
1996).  The results are plotted in Figure 1.  Given typical pulsar
ages of about 4 Gyr, there seems to be a reasonable agreement between
model and data in the sense that the median data values are close to
the theoretically predicted trend.  On the other hand, there is
considerable scatter about this median.  Furthermore, the 4 Gyrs
typical age is in line with what is expected from models that attempt
to explain the relative number densities of LMXBs and millisecond
pulsars seen in globular clusters today (e.g. Davies \& Hansen 1998).

The reasons for the scatter are likely to be many.  Some spread in
binary pulsar ages is likely; not all these pulsars are in the cores
of their host clusters; the effects of four-body interactions are not
considered by Heggie \& Rasio (1996); and some of the pulsars might be
in the tail of the space velocity distribution, especially if they
were formed in exchange interactions.

We note that there is a poor agreement for observed eccentricities
above about 0.01, but that this is likely to be due primarily to the
fact that this is the regime for which the power law relation between
eccentricity and waiting time changes into a logarithmic relationship.
Another clear outlier is the highly eccentric pulsar, M~15C, which is
also 13 core radii from the centre of the cluster, and is thought to
have been involved in a recent exchange interaction (Phinney \&
Sigurdsson 1991).  The cross-section for that interaction should have
been determined by an orbital period much longer than the current
period, and hence that the cross-section for the interaction which
produced the eccentric orbit may have been much larger than the
cross-section for the perturbation of its current orbit. In fact, its
predicted eccentricity is well below the range plotted here, while its
actual eccentricity is about 0.68, clearly indicating that some
mechanism other than ``gentle'' fly-by encounters produced its
eccentricity.  Another system, NGC 6752A and its low eccentricity
combined with its location so far out in the cluster have been used to
suggest that the system experienced a recoil from a black hole-black
hole binary in the core of NGC 6752 (Colpi, Possenti \& Gualandris
2002), again indicating that not all binary pulsars eccentricities
will be well described by the framework of gentle fly-by induced
eccentricities.  Still, though, as the bulk of the measurements and
the upper limits agree reasonably well with a $\sim4\times10^9$ year
pulsar lifetime and the basic model for inducing eccentricities of
Rasio \& Heggie (1995), it seems reasonable to use this approach as a
guide to estimating the number of moderately eccentric binaries there
will be in a globular cluster.

\begin{figure}
\centerline{\psfig{figure=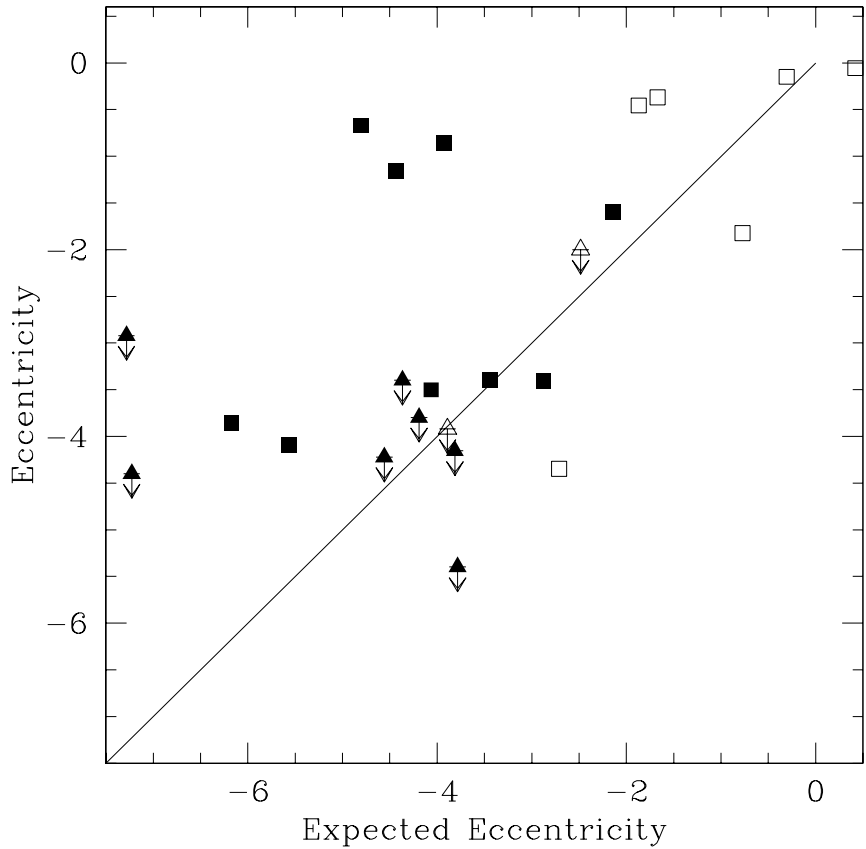,height=12 cm}}
\caption{The expected versus observed eccentricities for the globular
cluster pulsars tabulated by Camilo \& Rasio (2005).  Squares
represent observed eccentricity values, while triangles represent
upper limits.  The calculations assume a lifetime of the pulsar system
of $4\times10^9$ years, during which there has been no orbital
evolution except due to interactions with a third body.  The line
shows where the expected and observed eccentricities are equal, and is
plotted to guide the eye.  Filled symbols indicate pulsars whose
positions are well known, while open symbols indicate pulsars whose
positions are not well measured and which are hence assumed to be in
their cluster cores.}
\end{figure}

\subsection{The circularisation timescale}
The observational and theoretical status of the circularisation of
binaries containing giant donors is well established.  It is generally
found that the circularisation timescales of such systems is of order
10000 years (Zahn 1977; Verbunt \& Phinney 1995), with theory in good
agreement with the data.  The circularisation rate of short period
eccentric binaries containing low mass main sequence stars is not
currently well understood observationally or theoretically.
Observations of open clusters have done much to test theory for the
circularisation rates of longer period binaries, but it seems that
binaries with periods shorter than about 8 days are circularised
during their pre-main sequence phases of stellar evolution (Mathieu et
al. 1992).  As a result, if a shorter period binary forms after the
star has ended its pre-main sequence phase of evolution, there is not
a straightforward way to use observations of the tidal circularisation
cutoff periods of open clusters in order to estimate the
circularisation rate for such binaries.

We must, then rely on some combination of theoretical work and
extrapolation of the observational data.  The key question is thus
whether it is plausible for a few percent of the X-ray binaries to
have eccentricities greater than about 0.0001.  This essentially boils
down to the question of whether the tidal circularisation timescale
can be about 0.01 times the timescale on which perturbations produce
an eccentricity of about 0.0001.  In fact, a lower fraction could
still be consistent with the data, since the eccentric binaries have
their accretion rates boosted due to the eccentricity, and hence the
observable (i.e. flux-limited) sample of X-ray binaries should be
biased towards eccentric binaries.

The theory and observational understanding of circularisation of
eccentric binaries with low mass main sequence stars is still in its
early stages, due to uncertainties regarding convection and turbulence
within stars.  If one takes the most optimistic case in the
theoretical literature, that of Goodman \& Dickson (1998) in the case
of inefficient linear damping of tidally excited $g$-modes, one finds
that the circularisation timescale is about 150 $P_d^3$ Myrs, which
would be about 20 Myrs for the 0.5 day periods being discussed here,
and would give a few percent of the X-ray binaries in sufficiently
eccentric orbits to allow observable effects.  If one takes a more
conservative case, such as that of Witte \& Savonije (2002), then one
can extrapolate their published relation to estimate a circularisation
timescale of a few Myrs for periods of about 0.5 days, still
marginally consistent with the data.  Furthermore, the results of
Witte \& Savonije (2002) invoke resonant effects to speed up the
circularisation process.  These resonant effects are not likely to be
important for eccentricities below about 0.1, so the removal of the
last bit of eccentricity from a nearly circular binary would be
expected to happen on a slower timescale than a few Myrs.  On the
other hand, other groups, such as Terquem et al. (1998) and Claret \&
Cunha (1997) have suggested much more rapid circularisation of short
period binaries, with a stronger dependence of circularisation
timescale on binary period, and furthermore, the circularisation
process may be accelerated due to mass transfer in contact binaries.

The biggest problem most of the theoretical work has in matching the
observations is to explain the presence of the longest period circular
binaries; that is to say, most of the theoretical work seems to
underestimate the circularisation rate for one reason or another.  As
a result, it is probably safest to assume circularisation rates higher
than those typically proposed in the literature, at least for the
longer period binaries.  This still yields few constraints for the
circularisation timescales of the lower period binaries, since the
dependence of the circularisation timescale on period is not well
constrained.  At the present time, observations are ongoing to measure
circularisation periods in more open clusters (e.g. Meibom \& Mathieu
2005), but if the supposition that much of the circularisation of
normal binary stars occurs on the pre-main sequence is correct, then
such observational work will never address the question of
circularisation of short period binaries.  In fact, the globular
cluster X-ray binaries (along with other similar systems such as
binary pulsars in globular clusters) might then present the best test
ground for this work.  A large sample of flaring X-ray sources
associated with eccentric orbits through proof that the flares were
periodic could provide a key data set of great use in refining models
of tidal circularisation of short period binaries containing
solar-type stars.

\subsection{Observational tests}
The most obvious observational test of this model is that the flares
should be periodic, on the binary period.  Based on the fact that the
flares in the NGC~4697 sources were seen in multiple observations, we
infer that the orbital periods, at least of these sources, cannot be
much larger than the observation durations, which were about 11 hours.
The lack of multiple flares being seen in a single observation would
suggest that the periods are at least a bit longer than 11 hours,
although it was noted by Sivakoff et al. (2005) that the method of
detecting these flares relied on comparing the count rate in a
given window to the average count rate observed from the source, which
led to a bias against sources which show multiple flares.  Use of more
sophisticated statistical methods for flare finding such as Bayesian
blocks (Scargle 1998) might have greater sensitivity to shorter period
eccentric systems, which could show multiple flares.  Ideally, one
would have continuous observations of at least 30 hours on a single
galaxy which is at least as nearby as NGC~4697, or XMM observations,
with their higher count rates, on a galaxy at roughly the same
distance as NGC~4697.

Additionally, if larger samples of such flaring sources can be
collected, such flaring sources should be found preferentially in
dense globular clusters.  While some systems such as Circinus X-1 are
likely to exist, these should have short lifetimes after the supernova
explosion forming them (i.e. they should have be eccentric only for a
timescale of order the circularisation timescale).  As a result, these
should only exist in young stellar populations, even if they do have
low mass donors.  Therefore, elliptical galaxies should show such
periodically flaring X-ray binaries only when eccentricities can be
induced in the binary long after its formation.  This should happen
only in regions of high stellar density (i.e. globular clusters, and
to be more specifically, predominantly in the densest regions of the
densest clusters).

Additionally, we have some hope for finding evidence of these
eccentric accreting binaries in our own Galaxy, and using their
numbers to estimate how many such systems should exist in other
galaxies.  For example, the X-ray and optical properties of AC~211,
the accretion disk corona source in M~15, might be explained as the
result of the effects of binary eccentricity.  Van Zyl et al. (2004)
suggest that this system may have a rather large mass ratio, despite
its long orbital period.  This would imply a low mass transfer rate,
at odds with the inferred high mass transfer rate.  A large
eccentricity would boost the mass transfer rate substantially.
Obviously there are other ways to affect the mass transfer rate of
this system, and eccentricity is not currently required by the data.
A more effective way to test for eccentricity in this system would be
to obtain a more detailed radial velocity curve than currently exists.

It would also be fruitful to look at the eccentricity distributions of
the cataclysmic variable stars (CVs) in globular clusters in the Milky
Way.  As there are many more of these systems than millisecond
pulsars, it should be easier to collect a large sample of objects.
Unfortunately, at the present time, eccentricity measurements of
globular cluster CVs have not been made.  This is partly because only
recently have high resolution ultraviolet surveys started to unveil
the large populations of CVs in globular clusters (e.g. Knigge et
al. 2002), and partly because accurate measurements of very small
eccentricities are difficult to make except in pulsar systems.
Searches for eccentric CVs should become feasible in the near future
though, with adaptive optics spectroscopy, especially if CVs show
emission lines in the infrared, such as Bracket $\gamma$.  It might
also be possible to determine whether some magnetic CVs are in
eccentric orbits based on pulse timing.  Intermediate polar systems
show pulsation and orbital periods different from one another, and
using the pulsar as a clock, one can make accurate measurements of the
orbital parameters.

\section{Conclusions}
We have shown that the recent discovery of repeated flaring from two
of the X-ray binaries in NGC~4697 may be indicative of the fact that
these systems are in eccentric orbits.  The sharp, frequent increases
of the luminosity are consistent with the expectations for Roche lobe
overflow from a star in an eccentric orbit around a compact object.
The expected number of eccentric low mass X-ray binaries is consistent
with the observed number of systems showing repeating flaring
behaviour in NGC~4697.  We have also discussed future tests for
determining whether the assumptions about eccentricity evolution on
which this work is based are correct.

\section{Acknowledgements}
We are grateful for useful discussions with Ed van den Heuvel,
Christian Knigge, Rudy Wijnands, Simon Portegies Zwart, Alessia
Gualandris, Michael Sipior, Jasinta Dewi, Arjen van der Meer and
Gertjan Savonije and for comments from an anonymous referee which
helped to improve the clarity and focus of the paper.

\label{lastpage} 


\begin{thebibliography}{}
\bibitem{}Abramowicz M.A., Czery B., Lasota J.P., Szuszkiewicz E., 1988, ApJ, 332, 646
\bibitem{}Blaes O.M., 1987, MNRAS, 227, 975
\bibitem{}Camilo F., Rasio F.A., 2005, ASP Conf. Ser. Vol. 328: Binary Radio Pulsars, eds. F. A. Rasio \& I. H. Stairs (San Francisco: ASP), p. 147
\bibitem{}Claret A., Cunha N.C.S., 1997, A\&A, 318, 187
\bibitem{}Clark G.W., 1975, ApJ, 199L, 143
\bibitem{}Clarkson W.I., Charles, P.A., Onyett, N. 2004, MNRAS, 348, 458
\bibitem{}Colpi M., Possenti A., Gualandris A., 2002, ApJ, 570, 85L
\bibitem{}Cornelisse R., Heise J., Kuulkers E., Verbunt F., in 't Zand J.J.M., 2000, A\&A, 357, L21
\bibitem{}Cumming A., 2003, ApJ, 595, 1077
\bibitem{}Cumming A., Bildsten L., 2001, ApJ, 559, 127L
\bibitem{}Davies M.B., Hansen B.M.S., 1998, MNRAS, 301 15
\bibitem{}Fabian A.C., Pringle J.E., Rees M.J., 1975, 172P, 15 
\bibitem{}Fregeau J.M., Cheung P., Portegies Zwart S.F., Rasio F.A., 2004, MNRAS, 352, 1
\bibitem{}Gnedin O.Y., Zhao H.S., Pringle J.E., Fall S.M., Livio M., Meylan G., 2002, ApJ, 568, L23 
\bibitem{}Goodman J., Dickson E.S., 1998, ApJ, 507, 938
\bibitem{}Harris W.E., 1996, AJ, 112, 1487
\bibitem{}Heggie D.C., Rasio F.A., 1996, MNRAS, 282, 1064
\bibitem{}Hills J.G., 1976, MNRAS, 175P, 1 
\bibitem{}Hut P., Paczynski B., 1984, ApJ, 284, 685
\bibitem{}in 't Zand J.J.M., Kuulkers E., Verbunt F., Heise J., Cornelisse R., 2003, A\&A, 411, 487L
\bibitem{}Kaluzienski L.J., Holt S.S., Boldt E.A., Serlemitsos P.J., 1976, ApJ, 208, L71
\bibitem{}Knigge C., Zurek D.R., Shara M.M., Long K.S., 2002, ApJ, 579, 752
\bibitem{}K\"ording E., Falcke H., Markoff S., 2002,A\&A, 382L, 13
\bibitem{}Krolik J.H., Meiskin A., Joss P.C., 1984, ApJ, 282, 466
\bibitem{}Kundu A., Maccarone T.J., Zepf S.E., 2002, ApJ, 574 L5
\bibitem{}Kuulkers E., 2002, A\&A, 383, L5
\bibitem{}Kuulkers E., den Hartog P.R., in 't Zand J.J.M., Verbunt F.W.M., Harris W.E., Cocchi M., 2003, A\&A, 399, 663
\bibitem{}Kuulkers E., 2004, Nuclear Physics B., 132, 466
\bibitem{}Maccarone T.J., Kundu A., Zepf S.E., 2003, ApJ, 586, 814
\bibitem{}Mathieu R.D., Meibom S., Dolan C.J., 2004, ApJ, 602, 121L
\bibitem{}Mathieu R.D., Duquennoy A., Latham D.W., Mayor M., Mermilliod T., Mazeh J.C., 1992, Binaries as Tracers of Stellar Formation. Proceedings of a Workshop held in Bettmeralp, Switzerland, Sept. 1991, in honor of Dr. Roger Griffin. Editors, Antoine Duquennoy, Michel Mayor; Publisher, Cambridge University Press, Cambridge, England, New York, NY
\bibitem{}Meibom S., Mathieu R.D., 2005, ApJ, 620, 970
\bibitem{}Mirabel I.F., Rodriguez L.F., 1999,ARA\&A, 37, 409
\bibitem{}Phinney E.S., 1992, Phil. Trans. Roy. Soc. Lon., 341, 39
\bibitem{}Phinney E.S., Sigurdsson S., 1991, Nature, 349, 220
\bibitem{}Rappaport S., Putney A., Verbunt F., 1989, ApJ, 345, 210
\bibitem{}Rasio F.A., Heggie D.C., 1995, ApJ, 445L, 133
\bibitem{}Scargle J.D., 1998, ApJ, 504, 405
\bibitem{}Sivakoff G.R., Sarazin C.L., Jordan A., 2005, ApJ, 624L, 17
\bibitem{}Strohmayer T.E., Brown E.F., 2002, ApJ, 566, 1045
\bibitem{}Tauris T.M., Fender R.P., van den Heuvel E.P.J., Johnston H.M., Wu K., 1999, MNRAS, 310, 1165
\bibitem{}Terquem C., Papaloizou J.C.N., Nelson R.P., Lin, D.N.C., 1998, ApJ, 502, 788
\bibitem{}van Zyl L., Ioannou Z., Charles P.A., Naylor T., 2004, A\&A, 428, 935
\bibitem{}Verbunt F., Phinney E.S., 1995, A\&A, 296, 709
\bibitem{}Wijnands R., 2001, ApJ, 554, L59
\bibitem{}Witte M.G., Savonije G.J., 2002, A\&A, 386, 222 
\bibitem{}Zahn J-P., 1977, A\&A, 57, 383

\end{thebibliography}
\end{document}